\documentclass[a4paper,10pt]{article}
\usepackage{ucs}
\usepackage[utf8]{inputenc}
\usepackage{amsmath}
\usepackage{amsfonts}
\usepackage{amssymb}
\usepackage{amsthm}
\usepackage{rotating}
\usepackage{xcolor}
\usepackage[english]{babel}
\usepackage[T1]{fontenc}
\usepackage{graphicx}
\usepackage{hyperref}
\usepackage{cuted}
\usepackage{verbatim}
\usepackage{subfigure}
\usepackage{multirow}
\usepackage{bbold}
\usepackage{slashed}
\usepackage{indentfirst}
\usepackage{tcolorbox}
\usepackage{placeins}
\usepackage[a4paper, total={7in, 10in}]{geometry}
\usepackage{setspace} 
\newcounter{eq}

\newcommand{\bpsi}{\bar{\psi}}

\begin{document}


\title{\bf 
  $U_A(1)$ symmetry breaking quark 
interactions from vacuum polarization
} 

\author{ Fabio L. Braghin 
\\
{\normalsize Instituto de F\'\i sica, Federal University of Goias}
\\
{\normalsize Av. Esperan\c ca, s/n,
 74690-900, Goi\^ania, GO, Brazil }}

\date{}

\maketitle

\begin{abstract}
By considering the one loop background field method for 
 a quark-antiquark interaction, mediated by one (non perturbative) gluon exchange,
sixth order quark effective interactions are derived and investigated in the limit of zero momentum transfer for  large quark and/or gluon effective masses.
They extend fourth order quark   interactions worked out in previous works of the author.
These interactions break $U_A(1)$ symmetry and may be
either momentum  independent or dependent.
Part of these   interactions
vanish in the limit of massless quarks, and several other
- involving vector and/or axial quark currents - 
survive.
In the local limit of the resulting interactions, some phenomenological implications are
presented,  which  correspond to 
corrections to the Nambu-Jona-Lasinio model.
By means of the auxiliary field method, the local interactions give rise to 
three meson interactions  whose values are compared to phenomenological values found
in the literature.
Contributions for meson-mixing parameters
are calculated and compared to available results. 
\end{abstract}

\section{ Introduction}
 
The anomalous breaking of the $U_A(1)$ symmetry 
 was identified long ago 
   \cite{UA1,creutz-anomalies}
and it has been related to the non invariance of the 
measure of the functional integral 
in the Feynman path integral formalism \cite{fujikawa}.
This  symmetry breaking 
 has been shown to be responsible for 
different interesting effects in QCD and  hadron phenomenology.
In particular, it makes the
$\eta'$ meson to be considerably more massive than pions, the $\eta$
 and kaons
 \cite{thooft,witten}.
As a consequence, the pseudoscalar meson nonet 
becomes  similar to the vector nonet \cite{giacosa-etal-18}.
 Manifestations of properties due to QCD symmetries and possible corresponding 
symmetry breakings
 can also be, very often,
  analyzed by  means of  effective models 
whose  chiral and  flavor contents are directly identified to QCD 
\cite{witten-largeN,witten80,LSM,NJL,NJL2,pi-ren-2024,EffMod,several-Seff,pisarski-giacosa}.
In particular,  for  the Nambu-Jona-Lasinio (NJL) model,
the instanton induced  t' Hooft interaction is usually 
considered to implement $U_A(1)$ symmetry breaking   \cite{thooft,creutz-thooft}.
Several phenomenological consequences of this determinantal interaction
were found, mostly  related to meson  or flavor mixing
\cite{witten80,kawa+ohta,divecchia-etal,divecchia+vene,several-Seff}.
Accordingly,
't Hooft interaction makes possible 
the resolution of the  pseudoscalars $\eta-\eta'$ puzzle  
by means of the $\eta-\eta'$ mixing  
\cite{NJL2,creutz-thooft,hiller-etal-1}. 
The NJL model coupling constant has its grounds in one and two (dressed) 
gluon exchange.
Non-Abelian dynamics can be shown to produce an effective gluon mass for the 
gluon and, as such, it allows parameterizing the NJL coupling constant by 
$G\sim \alpha_s^2/M_G^2$ \cite{NJL-origin}.
It is interesting to note that  an interaction with the same shape as 
the 't Hooft  quark  effective  interaction in flavor SU(3) or U(3)
 NJL model,
 can also  be obtained from one loop
vacuum polarization, although it becomes different   for $N_f \neq 3$
 \cite{PRD-2014,PLB-2016,creutz-thooft}.
In the present work,  further one loop, 
$U_A(1)$ breaking, six-quark interactions
are  derived and investigated from a leading term of the QCD quark effective action. 
The behavior of the
 $U_A(1)$ symmetry breaking 
 with high energies/temperatures
has been debated for some time \cite{restoration,LQCD1,kovacs,EffMod}.
Some controversial  results in Lattice QCD (LQCD) were obtained
concerning if  its temperature restoration could  be 
close to the spontaneous chiral symmetry breaking restoration
 temperature \cite{restoration}
or if $U_A(1)$ symmetry would remain broken at higher temperatures
\cite{LQCD1,kovacs,pi-ren-2024}.  
Effects of the $U_A(1)$ interactions on the order of 
the chiral symmetry transition were also investigated \cite{UA1-CT1,UA1-CT2}.
By identifying further sources of $U_A(1)$ breaking,
 the investigation of 
its restoration must take into account the overall behavior in QCD,
not only the instanton induced processes.

The one loop  quark effective  action with
 background field 
constituent quark currents has been investigated in several works 
by considering both the Global Color Model (GCM) or the NJL-model
\cite{PRD-2014,PLB-2016,PRD-2022b,EPJA-2023,PRD-2021}.
Large quark mass 
 expansions of the determinant were performed, leading
to couplings of the  type of the NJL-model with extensions.
Similarly to the 't Hooft interaction,
 these polarization induced interactions   also disappear in
the   limit $N_c\to \infty$.
The use of auxiliary meson fields for quark-antiquark states
incorporates quark model  structure in terms of U(3)  flavor nonets
for 
pseudoscalar, scalar, vector and axial states.
The description of the light scalars, with their quark content,
 involves longstanding 
controversial results \cite{pelaez,JPG-2023} 
and they will not be really  discussed in the present work.
 The lightest axial  meson  multiplet (nonet)  is not unambiguous neither,
receiving further attention recently 
\cite{lightAxial}.
Axial meson production by different reactions
including two-vector meson fusion have 
been investigated and proposed in \cite{Lebiedowicz1}.
 Sixth order quark interactions lead to three-leg meson couplings 
that can contain
descriptions or contributions to meson (strong) decays 
that can be searched experimentally.
Some three-meson couplings of light mesons
were addressed in \cite{PRD-2022b}
and they will be revisited and extended in the present work by 
considering a different (re)normalization procedure.
The related physical  processes and mixings
can also contribute in a finite energy density environment
and contribute for the description of experimental results 
 on heavy ion collisions
\cite{rho-a1-1,rho-a1-2,rho-A1-mix-hic,pi0-eta}.
Several facilities plan to investigate related aspects,
that may involve specifically vector meson production and polarization,
FAIR/GSI, BESIII, J-PARC, JLAB, HERMES, SLAC,  NICA, LHC and EIC.

This work is organized as follows.
In the next section, the leading sixth order quark interactions are obtained from the
quark determinant by considering background quark currents.
These quark currents are dressed with a gluon cloud by means of components of an  
effective gluon propagator that can take into account non-Abelian gluon dynamics
\cite{PRD-2021,JPG-2022}.
These non-Abelian effects contribute to the strength of the quark-gluon interactions, such 
as to provide Dynamical Chiral Symmetry Breaking (DChSB).
As a consequence, it becomes useful to perform
a large quark effective mass that, in the present case, is basically equivalent to 
a large effective   gluon mass.
Four (constituent) color-singlet quark currents will be considered:
 scalar, pseudoscalar, vector and 
axial.
An effective gluon propagator
 will be chosen to make possible to perform numerical 
estimates for the meson dynamics.
The overall renormalization of the resulting terms will be provided by requiring that
the 4th order interactions correspond to the contributions to the NJL-model
coupling constant exactly as obtained in \cite{PRD-2021}.
In section (\ref{sec:consequences}) some phenomenological consequences 
are identified for the local limit of the interactions,  mainly in terms of the parameters of the NJL model.
For that, a mean field consideration for the scalar quark current is made 
such that   $J_{S,i}$ can be replaced by 
the quark chiral condensate for $i=0,3,8$ and zero for other components.
In section
(\ref{sec:mesondynamics})
all the local  interactions are rewritten in terms of local  auxiliary meson fields representing
the corresponding quark-antiquark mesons 
arranged in   U(3) flavor nonets. 
A (re)normalization condition is employed for this meson sector by
fixing each of the meson field normalization constants 
in the Lagrangian  kinetic terms that show up from the 
expansion of the same quark determinant.
Some consequences of the three-leg meson interactions
are extracted.
Meson mixing interactions   can  be defined.
In the last section there is a summary.

\section{ Sixth order quark interactions }
\label{sec:6interactions}

The basic steps of the whole calculation \cite{EPJA-2016,PLB-2016,PRD-2014} 
 are described below.
A low energy
 quark effective global color model
\cite{PRC,ERV}
  will be considered that is  given by:
\begin{eqnarray} \label{Seff}  
Z &=& N \int {\cal D}[\bpsi, \psi]
\;
\exp \;  i \int_x  \left[
\bar{\psi} \left( i \slashed{\partial} 
- m \right) \psi 
- 
 \frac{g^2}{2}\int_y j_{\mu}^b (x) 
({\tilde{R}}^{\mu \nu}_{bc} ) (x-y) j_{\nu}^{c} (y) 
\right]
,
\end{eqnarray}
Where 
$j^{\mu}_a = \bar{\psi} \lambda_a \gamma^{\mu} \psi$
is  the color  quark current, 
 the sums in color, flavor and Dirac indices are implicit, $\int_x$ 
stands for 
$\int d^4 x$,
$g^2$ the running
quark-gluon coupling constant, 
and  ${\tilde{R}}^{\mu \nu}_{bc}$ is a dressed gluon propagator.
To develop the flavor structure,
a Fierz transformation is performed and then  the background field method (BFM) is applied.
In the one loop BFM, it is enough to perform a shift in the 
quark currents, being  the  (quantum) quark current interactions  neglected and 
the quark field can be integrated out.
The interaction terms with mixed components of background currents (labeled by indices $_1$)
and quantum currents (indices $_2$) can be written as \cite{EPJA-2016}:
\begin{eqnarray}
\Omega_{12}  &\equiv&
  g^2 \left[
{J^S}_1  R {J^S}_2 + {J_i^P}_1  R {J_i^P}_2 
+ 
{J^S}_2   R {J^S}_1   
+ {J_i^P}_2  R {J_i^P}_1 
 \right]
\nonumber
\\
&-& 
\frac{ g^2}{2} \bar{R}_{\mu\nu} \left[  {J_i^\mu}_1   {J_i^\nu}_2  
+   {{J_i^\mu}_A}_1    {{J_i^\nu}_A}_2  
+ {J_i^\mu}_2     {J_i^\nu}_1  
+   {{J_i^\mu}_A}_2     {{J_i^\nu}_A}_1  
\right].
\end{eqnarray}
By using  $\lambda_i$ as the flavor Gell-Mann matrices,
with  indices of the adjoint representation  $i,j = 0...(N_f-1)^2$,
the non local   quark currents were defined respectively as
$ J_S^i  = J_S^i (x,y) =  (\bpsi \lambda^i \psi)$,
 $J_{PS}^i = J_{PS}^i (x,y) = (\bpsi i \gamma_5 \lambda^i \psi)$,
$J_{V,i}^\mu = J_{V,i}^\mu (x,y) =  (\bpsi  \gamma^\mu \lambda^i \psi)$ and
 $J_{A,i}^\mu  = J_{A,i}^\mu (x,y) = (\bpsi \gamma^\mu \gamma_5 \lambda^i \psi)$.
The functions $R = R(x-y)$ and $R^{\mu\nu} =  R^{\mu\nu}(x-y)$, carrying 
components of the gluon propagator,  can be written
 in momentum space in terms of 
  longitudinal and transversal components of the gluon propagator 
 as:  
\begin{eqnarray} \label{gluonpro}
  R(k) &=& 3 R_T(k) + R_L(k),
\nonumber
\\
 R^{\mu\nu}(k) &=& g^{\mu\nu} (R_T (k) + R_L (k) )
+ 2 \frac{k^\mu k^\nu}{k^2 } (R_T(k) - R_L(k) ).
\end{eqnarray}

The resulting quark determinant in the presence of  background constituent quark 
currents (indices $_1$ omitted from here on)  can be written as:
\begin{eqnarray} \label{Seff-det}  
S_{eff}   &=& - i  \; Tr  \; \ln \; \left\{
 i \left[ {S_0}^{-1}  +
\sum_\phi  a_\phi^i  J_\phi^i    \right]
 \right\} + I_4,
\end{eqnarray}
where 
$Tr$ stands for traces of all discrete internal indices 
and integration of  space-time coordinates,
$S_{0} = i / (i\slashed{\partial} - M )  $ is the quark propagator 
with either current quark masses ($m$) or 
constituent quark mass due to the DChSB ($M$).
  $I_4$ contains the background quark current   interactions - 
of fourth order in the background quark field -  that 
will not be considered further.
The background dressed quark currents,
for each Dirac channel $\phi = S, PS, V, A$,
appear  in chiral combinations with coefficients $a_\phi$ 
for  the selected channels given by:
\begin{eqnarray} \label{Rq-j}
a_S  
= 2 K_0 R  \lambda_i  ,
\;\;\;\;
a_{PS} = 
2 K_0 R  \lambda_i   \gamma_5,
\;\;\;\;
a_V =  - K_0
  {R}^{\mu\nu} 
  \lambda_i   \gamma_\nu ,
\;\;\;\;
\;\;\;\;
a_V =  - K_0
  {R}^{\mu\nu}  
  \lambda_i   \gamma_\nu \gamma_5.
\end{eqnarray}
where  $K_0 = 2 g^2/9$ being
$2/9$ from the Fierz transformation.
The background quark currents may contain $S,PS,V,A$ and flavor indices
either as super-indices or sub-indices  indistinctly, eg. $J_S^i = J_i^S$ and so on.

\begin{figure}[ht!]
\centering
\includegraphics[width=100mm]{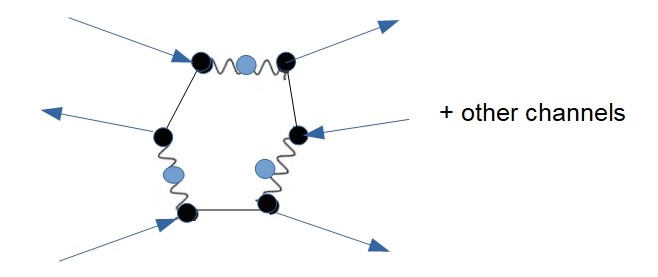}
\caption{ \label{fig:diagrams}
\small
Three quarks/anti-quark currents (solid lines)  interacting by 
components of a
dressed gluon  (wiggly line with a dot)
exchange
in one loop.
 }
\end{figure}

The leading one loop sixth quark/antiquark 
interaction
is shown in Fig. (\ref{fig:diagrams}),
where the wiggly lines with  circle represent 
components of the (dressed) effective gluon propagator as described above.
There are several ways to treat this determinant
and a direct large quark mass expansion
will be performed.

\FloatBarrier

The large quark mass expansion  goes along with a large effective gluon mass
expansion since all the quark currents are dressed by components of the gluon propagator,
and 
the third order terms of the expansion 
have the following shape:
\begin{eqnarray} \label{thirdO}
S^{(3)} \sim \frac{1}{3}
Tr \; 
\left[
\left( S_0 a_\phi J_\phi \right) \left( S_0 a_\phi J_\phi \right) \left( S_0 a_\phi J_\phi \right)
\right].
\end{eqnarray}
Next,   a derivative expansion can be applied along the lines
presented in \cite{mosel}
and
coefficients can be
resolved in momentum space
 for the  interaction terms defined in coordinate space  as powers of the currents.
For that, the quark propagator is written in the coordinate space so that 
the ordering in momentum 
 corresponding to the Feynman diagram can be done and
derivatives of quark currents may be considered.
The leading terms, described below, do not exhibit spacetime derivatives of 
quark currents.
The next leading terms contain one spacetime derivative acting in currents, and so on.
To resolve the couplings with the  zero order derivative expansion, one has the following general structure
for three arbitrary quark current for any channel $\phi_1,\phi_2,\phi_3$:
\begin{eqnarray}
S^{(3)} \sim   \frac{1}{3} \int_{x,y,z,..}
Tr_{F,D,C,k} \left( S_0 a_{\phi_1}  S_0 a_{\phi_2}  S_0 a_{\phi_3} \right) 
\; J_{\phi_1}(x,y)  \; J_{\phi_2} (y,z)\; J_{\phi_3} (z,x) 
+ ...,
\end{eqnarray}
where the trace $Tr_{F,D,C,k}$ is for flavor, Dirac, color indices and momentum
and $...$ stands for terms from the commutators that are non-leading.
Since the large quark mass /large gluon effective mass limit is considered for the expansion,
it is reasonable to restrict the quark currents to the local limit.
Therefore the leading second order terms of the expansion of the quark 
determinant, developed in \cite{PRD-2021,JPG-2022},
also do not involve quark currents derivatives.
They correspond to fourth order interactions for scalar-pseudoscalar currents,
being  treated in the zero momentum transfer and
local limits, so that they provide  corrections to the NJL coupling constant.
For the third order terms (sixth order interactions),
 there are many different structures involving the 
different quark currents.
Besides the leading  non-derivative quark current couplings, the next leading
single derivative couplings will be also considered below, being that 
further derivatives  are progressively  suppressed 
with respect to the non-derivative ones.
The first group of  leading terms for the flavor U(3)
contain terms with at least one scalar and/or pseudoscalar quark current,
either without any spacetime derivative or with one or two   derivatives.
The traces in Dirac and flavor indices select the resulting Lorentz scalar combinations of quark currents.
In the local limit, within zero momentum exchange,
the leading terms can be written as:
\begin{eqnarray} 
\label{6th-S-PS}
{\cal L}_{6,sb,S-PS} &=& 
-
 G_{sb,ps} 
 T^{ijk}  
( J^S_i  J^{PS}_j  J^{PS}_k + 
 J^{PS}_i  J^{PS}_j  S_k
+  J^{PS}_i  S_j  J^{PS}_k ) 
+
\; 
  G_{sb,s}   T^{ijk}  
 J^S_i   J^S_{j}  J^S_k
,
\\
\label{6th-SVV}
{\cal L}_{6,sb,SVV} &=&
 T^{ijk}  \left[ 
  3 \;G_{sb1} 
J^{V,\mu}_{i} J_{V,\mu}^{j}    J^S_{k}
+
3 \;     G_{sb2} 
  J^{A,\mu}_{i}  J_{A,\mu}^{j}
   J^S_{k}
+
3 i \; G_{sb1} \left(
J^{V,\mu}_{i}  J_{A,\mu}^{j}  J^{PS}_{k}  
-
  J^{A,\mu}_{i} J_{V,\mu}^{j}
J^{PS}_{k}
\right)
 \right],
\end{eqnarray}
\begin{eqnarray} 
\label{6d-SSV1}  
{\cal L}_{6d,SSV}
&=&
 3  \;     T^{ijk} 
\left\{ \left(  \partial_\mu J^{PS}_i \right) 
\left[ - G_{d1}
 J^S_{j}  
 J^{A,\mu}_{k}
+ G_{d2} 
 J^{A,\mu}_{j}
  J^S_{k}  
\right]
- G_{d2} (\partial_\mu J^S_i)
\left[
  J^{A,\mu}_j J^{PS}_k 
+  J^{PS}_j   J^{A,\mu}_k 
\right] 
\right.
\nonumber
\\
&+&
\left. 
(\partial_\mu A^{\mu}_{i}  )
\left[ 
G_{d1} J^S_j J^{PS}_k  + G_{d2} J^{PS}_j  J^S_k
\right]
\right\}
\nonumber
\\
&+& 
 3 \; i \;    T^{ijk}
\left[ 
(G_{d1} - G_{d2} )
J^{V,\mu}_i J^P_j
\left( 
\partial_\mu  P^{k}  \right) 
- 2 
 G_{d2} 
J^{V,\mu}_i
\left( 
\partial_\mu  P^{j}  \right)  J^P_k,
\right]
\nonumber
\\
\label{6-sva-sb} 
{\cal L}_{6,SVA,sb} &=& 
- 3 \;   G_{sva,sb}     T^{ijk}
\epsilon_{\mu\nu\rho\sigma}
\left\{ 2 
(\partial^\mu J^{V,\nu}_i )  (  \partial^\rho  J^{A,\sigma}_{j} ) J_S^k
+ (\partial^\mu J^{V,\nu}_i )    J_S^j (  \partial^\rho  J^{A,\sigma}_{k} )
\right\}
 + 
{\cal O } (\partial\partial ),
\end{eqnarray}
where the flavor structure was written
in terms of the symmetric and antisymmetric  SU(3) structure constants
$$T^{ijk} = 2 (d_{ijk} + i f_{ijk}),$$
 being that
the symmetric $d_{ijk}$ is directly extensible to $U(3)$.
In the last line 
$\epsilon_{\mu\nu\rho\sigma}$ is the Levi-Civita antisymmetric tensor.
The following flavor dependent coupling constants 
 were defined,  for  the limit of 
zero    momentum exchange:
\begin{eqnarray}
\label{G6s}
G_{sb,s}  T_{ijk} &=& 
 \frac{8 N_c}{3}  K_0^3  \;
Tr_D  \; Tr_F
\int_k
S_0 (k) \lambda_i   R(-k)  S_0 (k)  \lambda_j   R(k)
S_0 (k)  \lambda_k  R(-k) ,
\nonumber
\\
\label{G6ps}
G_{sb,ps} T_{ijk} &=& 
 \frac{8 N_c}{3} K_0^3  \;
Tr_D  \; Tr_F
\int_k
S_0 (k) i \gamma_5   \lambda_i   R(k) S_0 (k)  i \gamma_5   \lambda_j  
 R(k)
S_0 (k)  \lambda_k  R(-k),
\nonumber
\\
\label{G6SVV}
G_{sb1} T_{ijk} 
&=&  g^{\mu \rho}
 \frac{2 N_c}{3}  K_0^3  Tr_{D,F}
\int_k 
 {S}_0  (k) \gamma^\nu \lambda_i 
 R_{\mu\nu} (-k)
{S}_0  (k)
\gamma^\sigma \lambda_j  R_{\rho \sigma}  (-k)
 {S}_0  (k)
   \lambda_k
 R  (k) ,
\nonumber
\\
\label{G6PSVA} 
G_{sb2}  T_{ijk}
  &=& g^{\mu\rho}
\frac{2 N_c}{3}  K_0^3  Tr_{D,F}
\int_k 
{S}_0  (k)  \gamma^\nu  \gamma_5 \lambda_i   R_{\mu\nu} (-k)
{S}_0  (k)
\gamma^\sigma \gamma_5 \lambda_j R_{\rho \sigma}  (-k)
 {S}_0  (k) 
  \lambda_k
 R  (k) ,
\nonumber
\\
\label{dSSV}
G_{d1} T_{ijk}  &=& g^{\mu\rho}
\frac{4 N_c}{3}  K_0^3  Tr_{D,F}
\int_k 
 \tilde{S}_0  (k)  \gamma_\rho  \gamma^\nu \lambda_i   R_{\mu\nu} (-k)
 {S}_0  (k) \lambda_j  R  (-k)
 {S}_0  (k)
 \lambda_k
 R  (k) ,
\nonumber
\\
\label{dPSVAa}
G_{d2} T_{ijk} &=&  g^{\mu\rho}
\frac{4 N_c}{3}  K_0^3  Tr_{D,F}
\int_k 
{S}_0  (k)  \gamma^\nu \gamma_5 \lambda_i   R_{\mu\nu} (-k)
{S}_0  (k)
i\gamma_5 \lambda_j R  (-k) 
 \tilde{S}_0  (k) \gamma_\rho
 \lambda_k
R  (k) ,
\nonumber
\\
G_{sva,sb}  T_{ijk} &=&  - 
\frac{ \epsilon_{\rho_1 \mu \rho_2 \sigma}}{24}
\frac{2 N_c}{3} K_0^3 \; Tr_{D,F} \int_k 
\tilde{S}_0 \gamma^{\rho_1} \gamma_\nu \lambda_i R^{\mu\nu} (k) 
\tilde{S}_0 \gamma^{\rho_2} \gamma_\rho \lambda_j R^{\rho\sigma} (-k) 
{S}_0 i\gamma_5  \lambda_k R(-k) ,
\end{eqnarray}
where  $Tr_{D,F}$ stands for the traces in Dirac and Flavor indices,
$\tilde{S}_0 (k)  = 1/(k^2 - M^2)$ and $S_0(k)$ is the free quark propagator
with dressed masses.
Those coupling constants  with an index including $_{sb}$ 
are directly proportional to the quark mass, therefore corresponding to 
(chiral) symmetry breaking.
The effective constituent  quark mass will be considered to be constant,
and this lead to some simplification in the analysis.
All the other coupling constants,
  that do not include this index, 
are non-zero even in the case of massless quarks.
The last coupling ${\cal L}_{6,SVA,sb}$, with $G_{sva,sb}$, 
is an anomalous interaction
of a scalar current coupled  to vector and axial currents.

Further leading sixth order $U_A(1)$ symmetry breaking quark interactions
involve only vector and axial currents and at least one derivative, 
 that, in the local  limit for 
zero momentum exchange,  are given by:
\begin{eqnarray}
\label{L6dvva}
{\cal L}_{6v1}
&=&
3 \;   T^{ijk} \Gamma^{\rho\mu\alpha\beta}
\left[
-  G_{v1} \left( 
 (\partial_\rho J_{V,\mu}^{i}) J_{A,\alpha}^j J_{A,\beta}^k
+  (\partial_\rho J_{A,\alpha}^{i}) J_{A,\alpha}^j J_{V,\mu}^k \right)
-
G_{v2}  (\partial_\rho J_{A,\alpha}^{i}) J_{V,\mu}^j J_{A,\beta}^k
 \right]
\nonumber
\\
&+& 3 \; i \;  T_{ijk} 
\epsilon^{\lambda\nu\sigma\beta}
\left\{ G_{v1}
 (\partial_\lambda J_{V,\nu}^i ) 
J_{A,\sigma}^j J_{V,\beta}^k  
+
3  \; i \; 
 G_{v2}  
\left[
 (\partial_\lambda J_{V,\nu}^i ) 
J_{V,\sigma}^i  J_{A,\beta}^k 
-
 (\partial_\lambda J_{A,\nu}^i )
J_{V,\sigma}^j  J_{V,\beta}^k 
\right]
\right\}
\nonumber
\\
&-&
3 \; i \;  G_{v2}
T_{ijk} 
\; \Gamma^{\lambda\nu\sigma\beta} \;
 (\partial_\lambda J_{V,\nu}^i ) 
J_{V,\sigma}^j J_{V,\beta}^k  
\nonumber
+
3 \; i \;  G_{v1}
 T^{ijk}   E^{\lambda\nu\sigma\beta} 
(\partial_\lambda J_{A,\nu}^i )
A_{\sigma}^j J_{A,\beta}^k 
\end{eqnarray}
To write down a compact form for these coupling constants 
the gluon propagator was written in terms of a generic tensor
$R^{\mu\nu}(k) = \epsilon^{\mu\nu} \bar{R} (-k)$
for which $\epsilon^{\mu\nu} = g^{\mu\nu}$ for the longitudinal component
and $\epsilon^{\mu\nu} = (g^{\mu\nu} - k^\mu k^\nu/k^2)$ for the transversal one.
The following tensors were defined:
\begin{eqnarray}
T^{ijk} &=& 
 2 (d_{ijk} + i f_{ijk}),
\\
\Gamma^{\lambda \nu \sigma \beta}
&=& 
\left( g^{\lambda\mu}g^{\rho \alpha} 
- g^{\lambda\rho} g^{\mu\alpha} 
+ g^{\lambda\alpha} g^{\mu\rho} \right)
\epsilon_{\mu}^{\nu} \epsilon_{\rho}^{\sigma} \epsilon_{\alpha}^{\beta} ,
\\
E^{\lambda\nu\sigma\beta} &=& 
 \epsilon^{\lambda \mu \rho \alpha}
\epsilon_{\mu}^{\nu} \epsilon_{\rho}^{\sigma} \epsilon_{\alpha}^{\beta}.
\end{eqnarray}
 The coupling constants, calculated in the zero momentum exchange limit, 
limited for the case of transversal gluon propagators used below,
are the following:
\begin{eqnarray} \label{Gv1}
G_{v1} T_{ijk} \Gamma^{\lambda \nu \sigma \beta} &=& 
 \frac{2 N_c}{3}  K_0^3  Tr_{D,F}
\int_k 
\gamma^\lambda
\gamma_\mu \lambda_i 
\tilde{S}_0  (k) R^{\mu\nu} (-k)
\gamma_\rho \gamma_5 \lambda_j
 {S}_0  (k) R^{\rho \sigma}  (-k)
\gamma_\alpha  \lambda_k
 {S}_0  (k) R^{\alpha \beta}  (k) ,
\\
\label{Gv2}
G_{v2} T_{ijk} \Gamma^{\lambda \nu \sigma \beta} &=& 
 \frac{2 N_c}{3}  K_0^3  Tr_{D,F}
\int_k 
\gamma^\lambda
\gamma_\mu \gamma_5 \lambda_i 
\tilde{S}_0  (k) R^{\mu\nu} (-k)
\gamma_\rho  \lambda_j
 {S}_0  (k) R^{\rho \sigma}  (-k)
\gamma_\alpha  \lambda_k
 {S}_0  (k) R^{\alpha \beta}  (k) ,
\nonumber
\end{eqnarray}
These three coupling constants do not vanish in the limit 
of zero quark masses.
Also, 
 these coupling constants may have different values
for each of the flavor channel ($T_{ijk}$) if  quarks are non-degenerate
in the same way it has been done for the fourth order interactions \cite{PRD-2021,JPG-2022}.
In this case,
one must add the indices $^{(ijk)}$ 
  to each of the coupling constant.
The non zero momentum exchange case is straightforwardly obtained
either from the expansion of the determinant - provided all
the possible ordering of currents are considered.

\subsection{   The
effective gluon propagator and   renormalization condition
}
\label{sec:gluonprop+norm}

The effective gluon propagator to be considered in the integrals above may  
take into account a momentum dependent quark-gluon running coupling constant.
Nevertheless, it may be needed to incorporate the corresponding 
renormalization constants and corresponding parameters.
In the present work a transverse effective gluon propagator obtained
 from calculations with
Schwinger Dyson equations at the rainbow ladder level
in Refs.  \cite{gluonprop-SD} is used.
It is given by:
\begin{eqnarray} \label{gluonprop}
R_T (k) &=& 
\frac{8  \pi^2}{\omega^4} De^{-k^2/\omega^2}
+ \frac{8 \pi^2 \gamma_m E(k^2)}{ \ln
 \left[ \tau + ( 1 + k^2/\Lambda^2_{QCD} )^2 
\right]}
,
\end{eqnarray} 
with 
the following parameters 
$\gamma_m=12/(33-2N_f)$, $N_f=4$, $\Lambda_{QCD}=0.234$GeV,
$\tau=e^2-1$, $E(k^2)=[ 1- exp(-k^2/[4m_t^2])/k^2$, $m_t=0.5 GeV$,
$\omega = 0.5$GeV and $D= (0.55)^3/\omega$ (GeV$^2$).

All the consequences of the interactions above will be discussed 
for their local limit and zero momentum exchange
 so that those interactions will be seen as corrections to the NJL-model.
The gluon propagator normalization,  and the overall
constituent (dressed) quark current term  normalization,
was fixed such that the above one loop quark-antiquark interaction  
is the same as used in \cite{PRD-2021}.
This means that the 
{\it normalized} fourth order
 flavor-dependent corrections for the NJL -coupling constants, $G_{ij}$, 
 is reproduced for: 
\begin{eqnarray} \label{renorm-CQC}
\frac{G_{11}^{NJL} }{G_0} \simeq 1.0.
\end{eqnarray}

\section{  Some consequences}
\label{sec:consequences}

The resulting model obtained from the expansion of the determinant 
is non-renormalizable, containing 
interaction terms with  dimensionful coupling constants ($M^{-n}$ where $n$ is 
positive integer)
 that are UV finite.
Some coupling constants have dimension $M^{-5}$ (where $M$ is a mass 
dimension 
scale): 
$G_{sb,ps},  G_{6sb2}$ and 
$G_{sb,s} ,  G_{6sb1}$.
The  coupling constants of one single spacetime derivative terms
 have dimension $M^{-6}$: $G_{d1}, G_{d2}, G_{v1}$ and  $G_{v2}$.
The   $G_{sva,sb}$ coupling constant
 has  
dimension $M^{-7}$. 
The interactions of the second order (i.e. fourth order fermion interactions correcting
the NJL model) are also non-renormalizable.
Although these coupling constants are also UV finite, they've been employed
as corrections to the NJL-model that needs   UV cutoff scale to
produce observables, as a low energy effective model.
The numerical values of the coupling constants
 will be obtained with the normalization defined in the previous section \eqref{sec:gluonprop+norm}
in such a way to guarantee consistency.

Ratios of coupling constants provide a comparison of their relative 
order of magnitude
and they make possible to assess possible corresponding contributions to  observables.
There are basically  two types of  coupling constants from the point of view of their
momentum integrals.
For the present,  a constant quark effective mass is considered
what makes the analysis more transparent.
The following relations among the coupling constants arise
in the low energy regime for constituent quark masses instead of current masses
and degenerate quark masses:
\begin{eqnarray}
\frac{ G_{sb,ps} }{M}  &=& 
- \frac{G_{sb2}}{ M}
=  G_{d2} 
=  \frac{G_{v2} }{2},
\\
\frac{ G_{sb,s} }{M} &=&
  \frac{ G_{sb1} }{M} =
 G_{d1}
=
\frac{G_{v1} }{2}
\sim   4 M G_{sva,sb} ,
\end{eqnarray}
where the last relation of the second line is   
suitable for very large quark mass $M$ limit.
The one single derivative couplings (with $G_{d1/d2}$ and $G_{v1/v2}$)
are suppressed by a factor $1/M$ with respect to those
non-derivative couplings.
The anomalous coupling constant $G_{sva,sb}$ is suppressed further by $M^2$.

The couplings with  $G_{sb,ps/s} $ for scalar and pseudoscalar currents, 
in Eq. (\ref{6th-S-PS}) had already been derived in \cite{PRD-2014,PLB-2016}
for flavor SU(3) and SU(2)
in the limit of degenerate quark masses.
These flavor U(3) coupling has the same shape 
as those generated by the flavor U(3)  determinantal 't Hooft interaction 
\cite{thooft}.
However, for the case of 
any other  flavor group, $SU(N_f)$ or $U(N_f)$, 
the present method  still provides a sixth order interaction whereas
differently from the  determinantal
 't Hooft interaction \cite{creutz}.
Consequences of the $U_A(1)$ breaking  
 SU(3) 't Hooft interaction in hadron phenomenology
  have been investigated extensively  in NJL-type models
\cite{witten,NJL,NJL2,hiller-etal}.
In this type of models, these interactions
 drive mixing interactions for the  gap equations for the quark effective masses 
and for the bound state (Bethe-Salpeter) equations (BSE)
with results in good agreement with lattice QCD and phenomenology.
One may 
have an estimation  of their effects by resorting to a mean field approach
for the scalar current.
Given that the interactions were obtained by a large quark mass,
or alternatively by  a large gluon effective mass,
the local limit of the interactions may be considered without further conditions.
 In this case, one may
replace the scalar quark current by the scalar quark-antiquark condensate
$$
J_s^k  \sim < \psi \lambda^k \psi >.
$$
In Eqs. (\ref{6th-S-PS}) and (\ref{6th-SVV}), respectively,   this leads to 
different corrections to the NJL model and vector NJL-model coupling constants.
It can be written as:
\begin{eqnarray}
\label{6th-SVV-mf}
{\cal L}_{6,sb-NJL} &=&
\frac{G_{6,PS}^{ij}}{2}
  J_i^{PS} J_j^{PS} 
-  \frac{G_{6,S}^{ij}}{2}
J_i^S J_{j}^S  
+
\frac{G_{6,V}^{ij}}{2}
J^{\mu}_{V,i} J_{\mu}^{V,j}    
+
\frac{G_{6,A}^{ij}}{2}  J^{\mu}_{A,i} J_{\mu}^{A,j} ,
\end{eqnarray}
where 
\begin{eqnarray}
G_{6,S}^{ij} &=& 12 \; G_{sb,s}  d_{ijk}  < \bar{\psi} \lambda^k \psi > ,
\;\;\;\;\;
G_{6,PS}^{ij} = 12 \; G_{sb,ps} d_{ijk}  < \bar{\psi} \lambda^k \psi > ,
\nonumber
\\
G_{6,V}^{ij}  &=& 12 \; G_{sb1}   d_{ijk} < \bar{\psi} \lambda^k \psi > , \;\;\;\;\;\;
G_{6,A}^{ij} = 12 \; G_{sb2}    d_{ijk} < \bar{\psi} \lambda^k \psi >.
\end{eqnarray}
For the up, down and strange scalar quark-antiquark condensates, the components $k=0,3,8$
contribute as:
\begin{eqnarray}
< \bar{\psi} \lambda_0 \psi > &=& 
\sqrt{\frac{2}{3}}
\left( < \bar{u} u >   + < \bar{d} d >    +  < \bar{s} s > \right), 
\nonumber
\\
< \bar{\psi} \lambda_3 \psi > &=& 
  < \bar{u} u >  -  < \bar{d} d >     , 
\nonumber
\\
< \bar{\psi} \lambda_8 \psi > &=& 
\frac{1}{\sqrt{3}}
\left( < \bar{u} u >   + < \bar{d} d >    -  2  < \bar{s} s > \right).
\end{eqnarray}
These fourth-order quark-antiquark   corrections  break $U_A(1)$.
The resulting corrections for the NJL-model coupling constants must
be compared to an original coupling constant $G_0$
for   the normalization given in Eq. (\ref{renorm-CQC}).

 Numerical estimates for the 
 coupling constants  in Eqs. (\ref{G6s}) and
(\ref{Gv1})
are shown  in Table (\ref{table:Gcurrents})
by considering the parameters of  the fitting of the NJL given in \cite{PRD-2021}
for which  $G_0=10$ GeV$^{-2}$.
The following
phenomenological values of the 
 parameters:
\begin{eqnarray}
\label{parameters}
&& M_u = M_d =    M_s  =  0.391 \; \mbox{GeV}
\nonumber
\\
&& 
< \bar{u} u >  =  < \bar{d} d >=     (-0.240)^3   
\mbox{ GeV}^{3}, \;\;\;\;\;
  < \bar{s} s > =   (-0.290)^3 \; 
\mbox{ GeV}^{3}.
\end{eqnarray}
In few situations where mixing coupling constants are directly dependent 
on quark mass differences, the partially non-degenerate quark masses were employed:
 $M_u =M_d =  0.391$ GeV,      $M_s  =  0.600$ GeV.
Because of the ambiguities to define scalar and axial states to form quark-antiquark
 mesons nonets, numerical estimates will be provided mostly for
the up and down quark sector.

\begin{table}[ht]
\caption{
\small Numerical results for the resulting parameters for 
$T_{ijk} = 2 d_{11k}$ ($k=0,8$) , 
 by using parameters Eq. (\ref{parameters}) and effective gluon propagator
(\ref{gluonprop}).
For the mixing parameters (*) the following masses have been used:
 $M_u = M_d =  0.391$GeV,      $M_s  =  0.600$ GeV.
} 
\centering  
\begin{tabular}{| c  c c c c  | c  c c  | } 
\hline\hline  
  & $G_{sb,s} / G_{sb,ps}$  & $G_{sb1} / G_{sb2}$ &   $G_{d1} / G_{d2}$ 
& 
$G_{v1} / G_{v2}$  &  $G_{sva,sb}$ & 
$G_{6,S}^{11}$/$G_{6,PS}^{11}$
& $G_{6,V}^{11}$/$G_{6,A}^{11}$
  \\
\hline 
&  GeV$^{-5}$ &   GeV$^{-5}$   &
 GeV$^{-6}$ & 
 GeV$^{-6}$ &   GeV$^{-7}$ &  GeV$^{-2}$    & GeV$^{-2}$ 
\\ 
\hline 
$T_{118}$ &  2.9/86.7  & 2.9/-86.7  &59.3/221.7
& 118.7/443.5  &  -459.7  & -0.31/-9.4  &   -0.16/4.7
\\[1ex] 
\hline\hline  
& &   $G_{03}^{ps}$/$G_{03}^{s}$  & $G_{08}^{ps}$/$G_{08}^{s}$   
&   $G_{38}^{ps}$/ $G_{38}^{s}$  
&   $G_{03}^{v}$/$G_{03}^{a}$  & $G_{08}^{v}$/$G_{08}^{a}$ 
  &   $G_{38}^{v}$/$G_{38}^{a}$  
  \\
\hline 
&   &  GeV$^{-2}$ &   GeV$^{-2}$   &
 GeV$^{-2}$ &
 GeV$^{-2}$ &   GeV$^{-2}$  &    GeV$^{-2}$    
\\ [0.5ex]
\hline 
$T_{118}$  (*)&  &
 -0.08/-0.40  &  -0.79/-3.85   & -0.05/-0.23   &
-0.06/0.28 
& -0.56/2.72
 & - 0.03/0.16
\\[1ex] 
\hline  
\end{tabular}
\label{table:Gcurrents}  
\end{table}
\FloatBarrier

Note that, for  the instanton induced 't Hooft interaction one 
would have $G_{sb,s}  = 3 G_{sb,ps}$
whereas in the polarization induced coupling constant the values 
for the scalar and pseudoscalar  
channels are different.
By identifying the scalar part of the  't Hooft interaction 
(with coupling constant $\kappa$) 
with the 
scalar (or pseudoscalar) part $G_{sb,s/ps}$ one has: 
$
\kappa = - 32 \times G_{sb,s/ps}.
$
These values presented in the Table above  may therefore  smaller than,
or of the same order of magnitude as, 
different fittings
for the NJL model that are in the range of 
$ \kappa \sim - 100 \to  - 2500$ GeV$^{-5}$ \cite{NJL2,kohyama}.
It is important to notice that these 't Hooft interactions used in the NJL model
as a matter of fact can include the interactions from polarization processes,
at the end only a fitting for their numerical values is needed in such models.
Their (mean field) contributions for the flavor dependent 
NJL-coupling constants, 
 identified in Eq. (\ref{6th-SVV-mf}),
 are denoted 
in the Table
by $G_{6,S}^{11}/G_{6,PS}^{11}$ and their values are quite smaller than
$G_0=10$ GeV$^{-2}$.

In spite of the different nature, the coupling constants $G_{sb1}$ and $G_{sb2}$ are 
  vector and axial counterparts of 't Hooft interactions 
- being non-derivative.
 At the mean field level, analogously to the scalar-pseudoscalar terms,
they induce contributions for the vector-NJL coupling constants,
denoted by $G_{6,V}^{11}/G_{6,A}^{11}$ that are displayed
in the last  entry of the first line. 
They also are  much smaller than the typical values of the 
NJL model coupling constant.
Being proportional to the quark chiral condensate,  these 
coupling constants   go to zero with the 
restoration of DChSB.
Numerical values for the
 derivative interactions $G_{d1/d2}, G_{v1/v2}$ and $G_{sva,sb}$
in the channel $T^{118}$ are also displayed. Although
their numerical values are larger, they have smaller dimensions, 
respectively $M^{-6}$ and  $M^{-7}$.

Concerning the mixing parameters $G_{i \neq j}^{\phi}$, 
the values obtained  for the 
pseudoscalar/scalar mixing are of the same order of magnitude 
of those  obtained 
from the  second order terms of the  one loop polarization approach 
 \cite{PRD-2021} with a similar 
 normalization point to Eq. (\ref{renorm-CQC}).
 The same hierarchy of larger and smaller values are obtained,
in particular for the sets of  parameters $Y$ Table II of Ref.  \cite{PRD-2021}:
$$
G_{03}^{ps} \sim 0.03-0.11 \; (GeV^{-2}),
\;\;\;\;\;
G_{08}^{ps} \sim 1.04- 2.39 \; (GeV^{-2}) ,
\;\;\;\;\;
G_{38}^{ps} \sim 0.04-0.14 \; (GeV^{-2}).
$$
The mixing interactions for the  vector channels will be compared
to other works below for the
bosonized version of the interactions.

\section{ Corresponding mesons dynamics }
\label{sec:mesondynamics}

By associating each of the  flavor currents to a local meson field belonging to 
a flavor multiplet (nonet) - vector, axial, pseudoscalar and scalar-,
the above sixth order interactions
give rise to   three-meson couplings.
They can be associated to $U_A(1)$-breaking contributions to 
decay rates, although they may lead to rearrangements to meson mixing
or even  masses.
Local meson fields can be implemented by means of the auxiliary field method
with functional delta functions \cite{AFM-Z} as the following ones:
\begin{eqnarray}
1 &=&
N \int D [S_i, P_i ]  \delta (
  S_i  -  G_{0,s}^{ii}  J_i )   \delta (  P_i - G_{0,p}^{ii}  J_{P,i} )
\; 
\int D [V_\mu, A_\mu]  \delta  (  V^\mu_i  -  G_{0,v}^{ii}  J_{V,i}^\mu ) 
 \delta  (  A_i^\mu - G_{0,a}^{ii}  J_{A,i}^\mu )
,
\end{eqnarray}
where $N$ is a normalization,
$S_i, P_i, V_i^\mu, A_i^\mu$ are the  meson fields of each of the flavor multiplet,
$D [S_i, P_i ]$ and $D [V_\mu, A_\mu]$
are the measures of integration,
and $G_{0,s}^{ii}, G_{0,p}^{ii}, G_{0,v}^{ii}$ and $G_{0,a}^{ii}$  are  
constant parameters 
with dimension $M^{-2}$, to be fixed latter.
These last parameters may include the renormalization constants for each of the 
meson fields  implicitly what  is usually introduced in a more standard way 
with functional delta function such as the following:
$\delta (\phi \sqrt{Z}_\phi - G_0^{ii} J_{i}^\phi )$,
being that $\phi$ is any of the local meson fields and $G_0$ is a dimensionful constant
such as the NJL coupling constant \cite{JPG-2020b,renormaliz-AFM}.
Because the calculation will be basically done for degenerate quark masses,
the indices $^{ii}$ in $G_{0,\phi}$ will be suppressed in most part of the calculation
to make notation easier to read. 
 The fields $S_i$ will be referred as scalar  mesons fields
even if a mixing to other components (tetraquark, molecular or gluonic)
may be needed.

The local limit of 
the  Lagrangian terms above lead to the following contributions for the mesons dynamics:
\begin{eqnarray} 
\label{6th-S-PS-mes}
{\cal L}_{6,sb,S-PS}^{mes} &=& 
-
\frac{G_{sb,ps}}{
 G_{0,s} G_{0,ps}^2
} T^{ijk}  
( S_i P_j  P_k + 
 P_i  P_j  S_k
+  P_i  S_j  P_k ) 
+
\; 
\frac{ G_{sb,s}}{
 G_{0,s}^3
}  T^{ijk}  
S_i  S_{j} S_k
,
\\
\label{6th-SVV-mes}
{\cal L}_{6,sb,SVV}^{mes} &=&
 T^{ijk}  \left[ 
\frac{ 3 \;G_{sb1} }{
 G_{0,s} G_{0,v}^2
}  
V^{\mu}_{i} V_{\mu}^{j}    S_{k}
+
\frac{3 \;     G_{sb2} 
}{
 G_{0,s} G_{0,a}^2
}   A^{\mu}_{i} A_{\mu}^{j}
  S_{k}
+
  \frac{ 
3 i \; G_{sb1} \left(
V^{\mu}_{i} A_{\mu}^{j}  P_{k}  
-
 A^{\mu}_{i} V_{\mu}^{j}
 P_{k}
\right)  }{ 
 G_{0,v} G_{0,ps} G_{0,a}
}  
 \right],
\end{eqnarray}
\begin{eqnarray} 
\label{6d-SSV1-mes}  
{\cal L}_{6d,SSV}^{mes}
&=&
 3  \;   \frac{  T^{ijk}}{ G_{0,a} G_{0,s} G_{0,ps} }  
\left\{ \left(  \partial_\mu P_i \right) 
\left[ - G_{d1}
 S_{j}  
A^{\mu}_{k}
+ G_{d2} 
A^{\mu}_{j}
 S_{k}  
\right]
- G_{d2} (\partial_\mu S_i)
\left[
 A_j^\mu P_k 
+  P_j  A^\mu_k 
\right] 
\right.
\nonumber
\\
&+&
\left. 
(\partial_\mu A^{\mu}_{i}  )
\left[ 
G_{d1} S_j P_k  + G_{d2} P_j S_k
\right]
\right\}
\nonumber
\\
&+& 
 3 \; i \;  \frac{  T^{ijk}
}{  G_{0,v} G_{0,ps}^2 }   
\left[
(G_{d1} - G_{d2} )
V^\mu_i  
P_j
\left( 
\partial_\mu  P^{k}  \right) 
-
2 G_{d2}
V^\mu_i 
\left( 
\partial_\mu  P^{j}  \right) 
P_k
\right]
,
\nonumber
\\
\label{6-sva-sb-mes} 
{\cal L}_{6,SVA,sb}^{mes} &=& 
- 3 \; \frac{ G_{sva,sb}  }{  G_{0,a} G_{0,v} G_{0,s} }    T^{ijk}
E_{\mu\nu\rho\sigma}
\left\{ 2 
(\partial^\mu V^\nu_i )  (  \partial^\rho A_{j}^\sigma ) S^k
+ (\partial^\mu V^\nu_i )    S^j (  \partial^\rho A_{k}^\sigma )
\right\}
 + 
{\cal O } (\partial\partial ),
\end{eqnarray}
where some integrations by part have been done to simplify the results of the derivative terms.
\begin{eqnarray}
\label{L6dvva-mes}
{\cal L}_{6v1}^{mes}
&=&
3 \;  \frac{  T^{ijk} \Gamma^{\rho\mu\alpha\beta}
}{  G_{0,v} G_{0,a}^2 }   
\left[
- 2 G_{d2} \left( 
 (\partial_\rho V_{\mu}^{i}) A_\alpha^j A_\beta^k
+  (\partial_\rho A_\alpha^{i}) A_\alpha^j V_\mu^k \right)
-
G_{d1}  (\partial_\rho A_{\alpha}^{i}) V_\mu^j A_\beta^k
 \right]
\nonumber
\\
&+& 3 \; i \; \frac{T_{ijk} 
E^{\lambda\nu\sigma\beta}}{ G_{0,a} G_{0,v}^2
} 
\left\{ G_{v1}
 (\partial_\lambda V_{\nu}^i ) 
A_\sigma^j V_{\beta}^k  
+
3  \; i \; 
 G_{v2}  
\left[
 (\partial_\lambda V_{\nu}^i ) 
V_{\sigma}^i  A_{\beta}^k 
-
 (\partial_\lambda A_{\nu}^i )
V_{\sigma}^j  V_{\beta}^k 
\right]
\right\}
\nonumber
\\
&-&
3 \; i \; \frac{G_{v2}}{  G_{0,v}^3
}  T_{ijk} 
\; \Gamma^{\lambda\nu\sigma\beta} \;
 (\partial_\lambda V_{\nu}^i ) 
V_{\sigma}^j V_{\beta}^k  
\nonumber
+
3 \; i \; \frac{G_{v1}}{ G_{0,a}^3
}  T^{ijk}   \epsilon^{\lambda\nu\sigma\beta} 
(\partial_\lambda A_{\nu}^i )
A_{\sigma}^j  A_{\beta}^k 
\end{eqnarray}
For all of these couplings, renormalized coupling constants may be defined
by incorporating the constant factors $1/G_{0,\phi}$
and the numerical coefficients, such as 3 or 6.
Below, a criterion to determine these constants will be provided. 
These renormalized coupling constants 
can be written, for example, in 
the case of  degenerate quark mass as:
\begin{eqnarray}
G_{sb,ps}^R = \frac{G_{sb,ps}}{
G_{0,s}^3
},
\;\;\;\;
G_{sva,sb}^R =  3 \; \frac{ G_{sva,sb}  }{  G_{0,v}^2 G_{0,s} }
\;\;\;\;
G_{v1}^R = 3 \; \frac{G_{v1}}{ G_{0,v}^3}  \;, \;\;  ... 
\end{eqnarray}
The following types of three-meson 
 couplings  show up in the Lagrangian above:
the different scalar combinations of three scalar and pseudoscalar mesons  (SSS, SPP)
(\ref{6th-S-PS-mes}),
  vector-axial -pseudoscalar mesons (VPA) with 
$G_{sb1}$ and $G_{sb2}$  (\ref{6th-SVV-mes}), one scalar  
 with two vector or two axial mesons (SVV, SAA)
(\ref{6th-SVV-mes}),
 momentum dependent
 vector meson-two pseudoscalar mesons  (VPP),
or scalar-axial-pseudoscalar mesons - ASP, in  (\ref{6d-SSV1-mes}),
 momentum dependent combinations of three vector/axial mesons
(VVV,AAA) $G_{v1}, G_{v2}$ 
(\ref{L6dvva-mes}),
and   the anomalous (SVA)
$G_{sva,sb}$ (\ref{6-sva-sb-mes}).
There are also some VVA couplings with one single derivative with $G_{v1,v2}$.
 Higher order derivatives can also appear 
and  they 
may contribute to fusion of  two vector mesons into an axial meson,
even if they are non-leading, their energy dependence may be relevant.
The scalar current may represent a medium with finite baryonic density and, in the 
anomalous coupling $G_{sva,sb}$, it is 
needed to conserve linear momentum.
This interaction has been already presented in \cite{PRD-2022b}
and below it is treated within a different renormalization.
Other  interactions appear for higher order in the derivative whose coupling constants
are, however, still weaker.
The (derivative) vector and axial mesons couplings
may be part of (global) gauge invariant couplings  
as presented, for example,  in \cite{meissner} 
whose complete analysis is outside the scope of this work.
It will be enough to restrict the resulting 
terms from the expansion for the Abelian 
 stress tensors, that 
are  defined as $V_\mu = V_\mu^i \cdot \lambda^i$, by
$$
{\cal F}_{\mu\nu}^i = \partial_ \mu V_\nu^i - \partial_\nu V_\mu^i ,
\;\;\;\;\;\;\;
{\cal G}_{\mu\nu} = \partial_ \mu A_\nu^i - \partial_\nu A_\mu^i.
$$
Their complete contributions however will not be explored in the present work.

\subsection{ Parameters $G_{0,\phi}$: renormalization conditions }

The
meson kinetic terms, generated by the expansion of the determinant,
will be  used as renormalization conditions
to   determine the 
parameters $G_{0,\phi}$ for $\phi = S, P, V, A$ as proposed in \cite{JPG-2020b}.
The  free meson  kinetic terms   arise,
and they 
provide basically the full meson normalization constants.
The resulting meson kinetic terms, with the Abelian vector and axial mesons
stress tensors, are given by \cite{JPG-2020b,PRD-2022b}:
\begin{eqnarray}
{\cal L}_{mass} 
&=& 
\frac{1}{2} \frac{  I_{0,S}^{ii}}{ (G_{0,S}^{ii})^2 } 
 \partial_\mu S_i  \partial^\mu S_i 
+
\frac{1}{2} \frac{  I_{0,S}^{ii}}{ (G_{0,P}^{ii})^2 } 
 \partial_\mu P_i  \partial^\mu P_i 
-
\frac{1}{8} \frac{  I_{0,V}^{ii}}{(G_{0,v}^{ii})^2 }
 {\cal F}_i^{\mu\nu} {\cal F}_{\mu\nu}^i  
-
\frac{1}{8} \frac{  I_{0,A}^{ii}}{(G_{0,a}^{ii})^2 }
 {\cal G}_i^{\mu\nu} {\cal G}_{\mu\nu}^i  
\nonumber
\\
&=& 
\frac{1}{2} 
 \partial_\mu S_i  \partial^\mu S_i 
 + \frac{1}{2}  
 \partial_\mu P_i  \partial^\mu P_i 
-
\frac{1}{4}  
 {\cal F}_i^{\mu\nu} {\cal F}_{\mu\nu}^i  
-
\frac{1}{4}  
 {\cal G}_i^{\mu\nu} {\cal G}_{\mu\nu}^i  
\end{eqnarray}
Written in this way it is direct the association of the parameters $G_{0,\phi}$
to the  renormalization constants of the corresponding fields.
Therefore, it follows:
\begin{eqnarray} \label{G0-renorm}
G_{0,S}^i =    \sqrt{I_{0,S}^{ii}}
&=& 
G_{0,P}^i =  \sqrt{I_{0,P}^{ii}}
 ,
\nonumber
\\
G_{0,V}^i  =    \sqrt{I_{0,V}^{ii}/2} ,
&=& 
G_{0,A}^i =     \sqrt{I_{0,A}^{ii}/2}.
\end{eqnarray}
These parameters provide renormalized
expressions for the three-meson coupling constants shown above.

\subsection{
Estimation of $U_A(1)$ breaking contributions for the Mesons Masses 
and mixings}

 Meson masses can be computed  by solving
Bethe-Salpeter Equation (BSE) that may also be 
done  as usually done in NJL
or other approach \cite{NJL,NJL2,gluonprop-SD}.
By expanding the resulting Lagrangian terms that generate the corresponding 
BSE one can write down  effective Lagrangian terms for the masses of such mesons 
obtained from the BSE.
To provide an estimation, 
one can add to these effective Lagrangian terms, the
corrections obtained from the 6th order terms above.
By adopting a mean field value  for the scalar current as done above for the quark currents, 
it can be written as the scalar chiral condensate,
$\frac{ S_j }{ G_{0,s} } =  < \bar{\psi} \lambda_j \psi >$.
In this case, meson fields can be considered as parts of 
 U(3) flavor-nonets.
In this mean field approach one can write the corresponding effective contributions 
for the mass terms as:
\begin{eqnarray}
\label{MassUA1SPS}
\Delta {\cal L}_{S-PS}
&=&
- \frac{
(\Delta^{ps} M_{ji}^2 )
}{2}   P_j  P_k
- 
\frac{ 
 ( \Delta^{s} M_{ji}^2 ) 
}{2}S_j S_k
,
\;\;\;\;\;\; \mbox{from} \;\;\;\; {\cal L}_{6,sb,S-PS}.
\\
\label{MassUA1VA}
{\cal L}_{V-A} 
&=&
- \frac{ (\Delta^{V} M^2_{ji} )
}{2} 
V^{\mu}_{i} V_{\mu}^{j}     
-
\frac{ (\Delta^{A} M_{ji}^2 ) 
}{2} 
A^{\mu}_{i} A_{\mu}^{j},
\;\;\;\;\;\;
\mbox{from} \;\;\;\;
{\cal L}_{6,sb,SVV} ,
\end{eqnarray}
where  these mass corrections are the following 
\begin{eqnarray}
(\Delta^{ps} M_{ji}^2 )   &=& 
-
12 d_{jik}  \frac{{G}_{sb,ps}  }{G_{0,ps}^2}
< \bar{\psi} \lambda_k \psi >,
\;\;\;\;\;
(\Delta^{s} M_{ji}^2 )   = 
  12
 d_{jik}  \frac{{G}_{sb,s}  }{G_{0,s}^2}
< \bar{\psi} \lambda_k \psi >,
\nonumber
\\
(\Delta^{v} M_{ji}^2 )   &=& 
12  d_{jik}
\frac{{G}_{sb1}  }{ G_{0,v}^2 }
< \bar{\psi} \lambda_k \psi > ,
\;\;\;\;\;
(\Delta^{a} M_{ji}^2 )   =
12   d_{jik}
\frac{{G}_{sb2}  }{ G_{0,a}^2 }
< \bar{\psi} \lambda_k \psi > 
.
\end{eqnarray}
Some of these contributions will be  calculated below, namely the
ones for the lightest mesons of each multiplet:
the charged pion and kaon masses and corresponding 
scalar meson states $a_0(980), K_0^*(650-800$  respectively for 
$i=1,2$ and $i=4,5$
\cite{PDG,JPG-2022}, 
for the charged vector $\rho(770)$ and $K^*(892)$ masses
and corresponding axial mesons $A_1(1260) , K_1(1270)$ 
also  respectively for 
$i=1,2$ and $i=4,5$
\cite{PDG}.
The mass parameters $\Delta^\phi M_{ij}^2$ 
in Eqs. (\ref{MassUA1SPS}) and (\ref{MassUA1VA})  were written
in such a way to identify  also mixing terms for $i \neq j$.
The scalar/pseudoscalar mixing were compared to other available results above
at the level of quark currents.
For the vector channel, consider the
 possible mixings
between the components $\phi_0, \phi_3,\phi_8$ of each of the meson multiplet,
can be denoted by $G_{0,3}^{\phi}, G_{08}^{\phi}$ and $G_{3,8}^\phi$
can be written as:
\begin{eqnarray} \label{GmixV}
G_{0,3}^{R,v/a} \equiv  \;  -  \frac{ (\Delta^{v/a} M_{03}^2 )}{2}  &=&
- 12\;
 d_{033}  \frac{{G}_{sb,v/a}  }{G_{0,v/a}^2}
< \bar{\psi} \lambda_3 \psi >,
\nonumber
\\
G_{0,8}^{R,v/a} \equiv - 
 \frac{ (\Delta^{v/a} M_{08}^2 )}{2}  &=&
- 12 \; 
 d_{088}  \frac{{G}_{sb,v/a}  }{G_{0,v/a}^2}
< \bar{\psi} \lambda_8 \psi >,
\nonumber
\\
G_{3,8}^{R,v/a} \equiv 
- 
\frac{(\Delta^{v/a} M_{38}^2 ) }{2} &=&
- 12 \;
 d_{338}  \frac{{G}_{sb,v/a}  }{G_{0,v/a}^2}
< \bar{\psi} \lambda_3 \psi > .
\end{eqnarray}

In each of the  mesons flavor U(3) multiplets under analysis
\cite{weinberg},
pseudoscalar, scalar, vector and axial channels,
the following neutral meson mixings arise:
$$
\pi^0-\eta-\eta', 
\;\;\;\;\;
\rho^0-\omega-\phi,
\;\;\;\;\;
a_1^0(1260)-f_1(1280)-f_{1S}(1420)
$$
and the controversial case of the light scalars 
for which the quark-antiquark components may contribute for 
$\sigma(500)-a_0^0(980)-f_0(980)$ \cite{PDG,pelaez,JPG-2023}
In the pseudoscalar sector, 
a mixing coupling between pion-kaon, in such mean field for eqs. (\ref{MassUA1SPS})
with $G_{sb,ps}^{ijk}$,
 is forbidden since it would not conserve strangeness or isospin.
This is seen in the absence of symmetric structure constants
that relate $i=1,2,3$ and $i=4,5,6,7$.
 Results for the three-leg coupling constants and contributions for 
masses and mixing parameters
 are presented in the Table (\ref{table:meson-sector}).

\begin{table}[ht]
\caption{
\small Some numerical results for the  
(renormalized) parameters in the meson sector 
 by using parameters Eq. (\ref{parameters}) and  effective gluon propagator
(\ref{gluonprop}).  
For the mixing parameters (*) the following masses have been used:
 $M_u =0.35$GeV, $M_d =  0.355$GeV,      $M_s  =  0.65$ GeV.
} 
\centering 
\begin{tabular}{|c c  c c c  |c   c c   c | }  
\hline\hline   
&  $G_{sb,ps}^R$  &  $G_{sb,s}^R$   &   $G_{sb1}^R$  
&  $G_{sb2}^R$    & 
$G_{d1}^R$  & 
$G_{d2}^R$   & 
 $G_{v1}^R$ &  $G_{v2}^R$
  \\
\hline 
& GeV &  GeV  &   GeV    &
 GeV  &   
-  &   -  &  -  & - 
\\ [0.5ex]
$T_{118}$ & 
0.013 & 0.004  & 0.019 & 0.071 & 0.182  & 0.049
&  0.514  & 0.138 
\\[1ex] 
\hline\hline  
&   $\Delta M_{\pi}^2 $  & $\Delta M_{K}^2 $   &   $\Delta M_{A_0}^2 $  
& $\Delta M_{K_0^*}^2$  &  $\Delta M_{\rho^\pm}^2$
&$\Delta M_{{K^*}^\pm}^2$  &  $\Delta M_{A_1^\pm}^2$
 & $\Delta M_{K_1^\pm}^2$
  \\
&   (i=1,2)  & (i= 4,5)   &   (i=1,2)  & (i= 4,5)  
  &   (i=1,2)  & (i= 4,5)    &   (i=1,2)  & (i= 4,5)  
  \\
\hline 
 & GeV$^2$ &   GeV$^2$   &
 GeV$^2$ &   GeV$^2$ &  
 GeV$^2$ &   GeV$^2$  &  GeV$^2$ &   GeV$^2$ 
\\ [0.5ex]
 & $ 9.9\times 10^{-4}$ & $-12.6 \times 10^{-4}$   &
 $3.7\times 10^{-3}$ 
 & $-4.7\times 10^{-3}$ &
0.004   &  - 0.005
   & 0.015
 & -0.019   
\\[1ex] 
\hline\hline  
& &
  $G_{03}^{R,ps}$/$G_{03}^{R,s}$  & $G_{08}^{R,ps}$/$G_{08}^{R,s}$   
&   $G_{38}^{R,ps}$/ $G_{38}^{R,s}$  
&   $G_{03}^{R,v}$/$G_{03}^{R,a}$  & $G_{08}^{R,v}$/$G_{08}^{R,a}$ 
  &   $G_{38}^{R,v}$/$G_{38}^{R,a}$   & 
  \\
\hline 
& &  GeV$^{2}$ &   GeV$^{2}$   &
 GeV$^{2}$ &
 GeV$^{2}$ &   GeV$^{2}$  &    GeV$^{2}$     & 
\\ [0.5ex]
 (*) &  &  -0.06/-1.78 &  -0.58/-17.33   & -0.03/-1.03   &
-0.06/1.78 
& -0.58/17.33
 & - 0.02/1.03 & 
\\[1ex] 
\hline 
\end{tabular}
\label{table:meson-sector} 
\end{table}
\FloatBarrier

The couplings $SSS$ and $SPP$ correspond to similar structures
 of the bosonized 't Hooft interaction as discussed above, for which 
the following (flavor-independent) relation  holds:
$G_{sb,ps}^{tHooft} = 3 G_{sb,s}^{tHooft}$, that is different from the obtained above.

The coupling $V-P-A$ in ${\cal L}_{sb,SVV}^{mes}$ with $G_{sb1/2}^R$
corresponds to vector-pseudoscalar-axial mesons vertex whose phenomenological 
values in the literature are  one   order of magnitude larger than the ones present in the Table.
For example, in Ref. \cite{du-zhao} 
the vertex $\pi-\rho-A_1$
was found to be:
$G_{v-a-\pi}   \sim 2 M \sim  0.7  \; \mbox{GeV}$.
A similar estimate   to the present work  was done in
\cite{PRD-2022b},  for a different (re)normalization, that is
 $G_{A\pi V} \sim 0.7- 1.2$.
 Also from \cite{du-zhao} VPA coupling was adopted to provide decay and production of axial mesons
with $G_{VPA} = 2.04 $GeV that is   larger than values obtained above.
The coupling constants  $G_{sb1/2}^R$ also drives couplings of the type
$V-V-S$ and $A-A-S$ discussed above.

 The couplings  $V-P-P$ in ${\cal L}_{sb,SVV}^{mes}$ with $G_{d1/d2}^R$
stand for vector meson - two pseudoscalar mesons.
These coupling constants also drive 
 axial-pseudoscalar-scalar  mesons vertices ($A-S-P$).
In \cite{du-zhao} different values were considered for the different channels ASP, either 
real or complex, with moduli of the order of $1-3$ GeV 
that are somewhat larger than the values for $G_{d1,2}$ presented in the Table.
For the lightest quark sector it corresponds to the vertex $\rho-\pi-\pi$
such that 
$G_{\rho-\pi-\pi} = 3 (G_{d1}^R \pm  G_{d2}^R)$.
For this coupling constant  there are several estimates in the literature
that are  of the order of magnitude of  the values presented 
in the Table (\ref{table:meson-sector}).
 For instance,  from \cite{PDG}:
$G_{\rho-\pi\pi} = \sqrt{ 4 \; \pi \; 2.9 } \sim 6$.
A range of values for different   parameters cutoffs
 were provided  from other theoretical and experimental analysis for axial meson decay 
 for
$g_{VVA} \sim 9.27\to 35.02 $ and $5.30 \to 20.03$
\cite{Lebiedowicz1} that are large in comparison with $G_{v1/2}$.

The numerical values of  the  quantities  $\Delta M_\phi^2$ 
provide a naif estimation of the effects of the corresponding interactions
 on the meson mass.
Although
meson masses are obtained by solving the corresponding BSE,
these terms may be taken as corrections to the 
effective action with the physical  meson masses previously calculated 
with the corresponding BSE. 
These BSE are the ones previously solved with the same parameters considered 
in this work from \cite{PRD-2021}.
However, the relative magnitude and the corresponding sign can indicate
which mesons would receive larger contributions from this term, 
with the relative sign that may be positive or negative.
For the charged meson states built up with flavor generators $\lambda_1, \lambda_2$ 
 ($\lambda_4, \lambda_5$, therefore with one strange quark or antiquark)
the contribution of this quadratic term is positive  (negative).

The scalar-pseudoscalar meson mixings 
($G_{i \neq j}^{R,ps}/ G_{i \neq j}^{R,s}$)
are of the same order of magnitude of those
derived and  investigated  in \cite{PRD-2021,JPG-2022,JPG-2023},
although  a different procedure for the overall normalization have been considered.
  The vector mesons mixings ($G_{i \neq j}^{R,v}$)
are one order of magnitude larger than 
results presented in the literature   \cite{omega-phi}.
However, the hierarchy is the same, i.e.
the rho-omega mixing ($G_{03}^R$) is one order of magnitude 
smaller than the omega-phi mixing $G_{08}^R$).
The rho-phi mixing ($G_{38}^R$) is the smallest one.
The axial meson mixings ($G_{i \neq j}^{R,a}$)
are not well established, being these mesons very unstable.

Several of the three mesons couplings, shown above,
in the adopted mean field approach,
 represent corrections to 
 not only the controversial 
 so-called  pion- axial meson mixing,  $\pi -A_1$, coupling
\cite{osipov-etal},  but also 
correspondingly kaon-axial meson $K-K_1$ mixing and others,
 being all momentum dependent 
couplings.  
These couplings were already presented 
for the fourth order terms of the quark determinant expansion
in \cite{PRD-2022b}.
By adopting the  mean field approach  for
the scalar currents,  the following mixings are found:
\begin{eqnarray} 
\label{A-V-mixing}
{\cal L}_{6d,SSV-mes}
&=&
G_{A-P}^{R,(ij)}
\left[ (\partial_\mu A^{\mu}_{i} )
 P_j
-
A^{\mu}_{j} \left(  \partial_\mu P_i \right)
\right] ,
\;\;\;\;\;\;\;\;
G_{A-P}^{R,(ij)} =  - 6 \;   d^{ijk}   \frac{G_{d1} }{ G_{0,a}   G_{0,ps} } 
<\bar{\psi} \lambda^k \psi > ,
\end{eqnarray}
For the case of the lighter mesons 
$ij=1,2,3$,  pions and $A_1$,
one has the values of the Table for $G_{d1},G_{d2}$.
The following value is found:
\begin{eqnarray}
\label{GPA+GVA}
G_{A_1-\pi}^R, \simeq 0.17  \; \mbox{GeV}.
\end{eqnarray}
The $\pi-A_1$ mixing is usually eliminated by a redefinition of the axial meson 
field such as 
$A_\mu \to A_\mu + K \partial_\mu P$.
If a field redefinition of the $A_1$ meson field is performed to 
eliminate such mixing in the second order mixing term \cite{morais-etal}, 
the mixings above   (\ref{A-V-mixing}) - of third order -  still survive.
So the mixing should have to be done order by order in the determinant expansion.
It seems that this coupling  will not be eliminated consistently in all orders by a single field redefinition
of the A1.
However, a complete proof did not seem 
possible so far.
Otherwise, by trying to do the field redefinition in the quark determinant
before expansion, the derivative couplings of the pion would disappear.

The  third order correction to the anomalous
coupling $G_{sva,sb}^{R}$ 
can be associated to a highly momentum dependent 
 vector meson-axial meson mixing in the presence
of a quark scalar current,  scalar meson  or  finite density medium.
As discussed in \cite{PRD-2022b}, for the second order coupling,
this anomalous vector-axial mixing
 it is strongly dependent on the mesons polarizations
and it can only take place close to a third particle or in a medium to make 
possible the conservation of momentum, since axial and vector mesons
momenta must be transverse to each other.
It has been argued in that work that it makes possible to probe the 
axial constituent quark (baryon) current or charge by means of 
vector mesons.
It can be written for the mean field level:
\begin{eqnarray}
\label{6-sva-sb-1mes}
{\cal L}_{6,SVA,sb} &=& 
G_{A-V}^{R,(ij)}  \;  
\epsilon_{\mu\nu\rho\sigma} 
(\partial^\mu V^\nu_i )  (  \partial^\rho A_{j}^\sigma )  ,
\;\;\;\;\;\;\; 
G_{A-V}^{R,(ij)} = 
8 d^{ijk}  \; \frac{ G_{sva,sb}  }{  G_{0,a} G_{0,v}  }  <\bar{\psi} \lambda^k \psi >,
\end{eqnarray}
For the case of the lighter mesons 
$ij=1,2,3$ for the $\rho^0$ and axial $A_1$
one has the values of the Table
that yields
\begin{eqnarray}
\label{GPA+GVA}
 G_{A_1-\rho}^{R,(11)} \simeq  -0.07.
\end{eqnarray}
This (dimensionless) value is  quite smaller than the others and  it must be difficult to 
be observed experimentally because the axial mesons are 
quite unstable and the final state is  strongly interacting.
This type of mixing has been considered in many situations
for finite (energy/baryonic) densities including in heavy ion collisions
\cite{rho-a1-1,rho-a1-2,leopold} 
being therefore proportional to the baryonic density as 
$
\xi_{\rho-A_1} = C_{V-A} (\rho_N)/\rho_0  
$
where $C_{V-A} \sim 0.3$.

\section{ Summary }

 Sixth order $U_A(1)-$ breaking  quark interactions  were 
obtained  in the large quark mass expansion of the  quark determinant in the 
presence of background (constituent) quark currents.
The corresponding local limit was analyzed.
Four types of quark currents were considered:
scalar, pseudoscalar, vector and axial ones.
For the case of  non-degenerate quark masses, all the coupling constants,
as the limit of zero momentum transfer,
break flavor symmetry.
These  resulting  effective 
interactions are of the order of $\alpha^{2n}_s$, $n=3$,
where $\alpha_s$ is the quark-gluon running coupling constants.
In   energies lower than the dynamical chiral symmetry restoration,
 the effective  coupling constants depend on the  quark effective masses.
For energies around  and above the restoration
 of Dynamical Chiral Symmetry Breaking  \cite{ChRest},
besides thermal effects, 
the running quark-gluon coupling constant is considerably weaker
 and
 quark masses   reduce to the current quark masses
\cite{PDG}, although they  should not vanish  
 \cite{creutz-anomalies}.
Therefore, the strength of such  sixth order quark-antiquark 
 interactions  may decrease drastically, in particular 
 for those interactions proportional  to  quark masses, labeled by $_{sb}$.
 However,
they may not disappear,
 in agreement with 
\cite{kovacs,pi-ren-2024}.
By resorting to a mean field approach to the scalar quark currents
in the local limit of the interactions,
several sixth order interactions reduce to fourth order interactions
which
make possible to exploit
some phenomenological consequences    
in terms of the parameters of a NJL-model with the corresponding dynamics.

 By employing the auxiliary field method,
   the quark-current interactions were rewritten as
  three-leg meson vertices that were investigated  in the local  limit.
The  limit of 
zero momentum exchange adopted in these estimations provide an upper bound for 
the resulting three-meson running coupling constants.
Some coupling constants were estimated and  compared to other works, showing a 
reasonably good agreement
for the order of magnitude.
All the vertices contain the overall structure  from both 
the symmetric and the asymmetric flavor  structure constants, although
in some cases only one or the other may contribute.
Although these three-meson leg vertices are associated to meson decays,
 the investigation of decay rates were left outside the scope of this work and 
will be presented in a forthcoming work.

 \vspace{0.5cm}

\centerline{\bf Acknowledgements}

The author thanks a  short conversation with 
T. Kov\'acs. 
F.L.B. is a member of
INCT-FNA,  Proc. 464898/2014-5
and  he acknowledges partial support from 
 CNPq-312750/2021 and CNPq-407162/2023-2.

\end{document}